\documentclass{article}

\usepackage{arxiv}

\usepackage[utf8]{inputenc} 
\usepackage[T1]{fontenc}    
\usepackage{lmodern}        
\usepackage{hyperref}       
\usepackage{url}            
\usepackage{booktabs}       
\usepackage{amsfonts}       
\usepackage{nicefrac}       
\usepackage{microtype}      
\usepackage{graphicx}

\title{Simulation of Survival Data With the Package \pkg{rsurv}}

\author{
    Fábio N. Demarqui
   \\
    Department of Statistics \\
    Universidade Federal de Minas Gerais \\
  Belo Horizonte, Brazil \\
  \texttt{\href{mailto:fndemarqui@est.ufmg.br}{\nolinkurl{fndemarqui@est.ufmg.br}}} \\
  }

\usepackage{color}
\usepackage{fancyvrb}

\DefineVerbatimEnvironment{Highlighting}{Verbatim}{commandchars=\\\{\}}
\usepackage{framed}
\definecolor{shadecolor}{RGB}{248,248,248}
\newenvironment{Shaded}{\begin{snugshade}}{\end{snugshade}}

\newcommand{\AttributeTok}[1]{\textcolor[rgb]{0.13,0.29,0.53}{#1}}

\newcommand{\CommentTok}[1]{\textcolor[rgb]{0.56,0.35,0.01}{\textit{#1}}}

\newcommand{\ConstantTok}[1]{\textcolor[rgb]{0.56,0.35,0.01}{#1}}
\newcommand{\ControlFlowTok}[1]{\textcolor[rgb]{0.13,0.29,0.53}{\textbf{#1}}}

\newcommand{\DecValTok}[1]{\textcolor[rgb]{0.00,0.00,0.81}{#1}}

\newcommand{\FloatTok}[1]{\textcolor[rgb]{0.00,0.00,0.81}{#1}}
\newcommand{\FunctionTok}[1]{\textcolor[rgb]{0.13,0.29,0.53}{\textbf{#1}}}

\newcommand{\NormalTok}[1]{#1}

\newcommand{\OtherTok}[1]{\textcolor[rgb]{0.56,0.35,0.01}{#1}}

\newcommand{\SpecialCharTok}[1]{\textcolor[rgb]{0.81,0.36,0.00}{\textbf{#1}}}

\newcommand{\StringTok}[1]{\textcolor[rgb]{0.31,0.60,0.02}{#1}}

\begin{document}
\maketitle

\begin{abstract}
In this paper we propose a novel \proglang{R} package, called
\pkg{rsurv}, developed for general survival data simulation purposes.
The package is built under a new approach to simulate survival data that
depends deeply on the use of \pkg{dplyr} verbs. The proposed package
allows simulations of survival data from a wide range of regression
models, including accelerated failure time (AFT), proportional hazards
(PH), proportional odds (PO), accelerated hazard (AH), Yang and Prentice
(YP), and extended hazard (EH) models. The package \pkg{rsurv} also
stands out by its ability to generate survival data from an unlimited
number of baseline distributions provided that an implementation of the
quantile function of the chosen baseline distribution is available in
\proglang{R}. Another nice feature of the package \pkg{rsurv} lies in
the fact that linear predictors are specified using \proglang{R}
formulas, facilitating the inclusion of categorical variables,
interaction terms and offset variables. The functions implemented in the
package \pkg{rsurv} can also be employed to simulate survival data with
more complex structures, such as survival data with different types of
censoring mechanisms, survival data with cure fraction, survival data
with random effects (frailties), multivarite survival data, and
competing risks survival data.
\end{abstract}


\section{Introduction}\label{introduction}

Survival analysis is a branch of statistics concerned with analyzing the
time taken from an event of interest to occur. The main characteristic
distinguishing survival analysis from other statistical techniques is
the existence of incomplete observations, known as censored
observations. The presence of censoring requires specialized methods for
the analysis and simulation of survival data.

There are different types of censoring. We say the survival data is
right-censored when the time until the event of interest is greater than
the observed time. Similarly, when the observed time is greater than the
time until the event of interest, we have left-censored survival data.
Finally, there are situations where the time to the event of interest is
only known to lie in an interval \([L, R]\). In this case, we say the
survival data is interval-censored. It is also important to know the
mechanism type that generates censoring when modeling and simulating
survival data. If the survival and censoring times are independent, we
have what is called independent censoring. In contrast, when the
survival and censoring times are not independent, we have a dependent
censoring mechanism. Under dependent censoring, a joint distribution for
the survival and censoring times needs to be specified, posing
additional challenges in the modeling and simulation process.

The large number of existing survival models today, combined with the
different types of censoring, makes the simulation of survival data a
challenging task. Unfortunately, by the time of writing, there are only
a few \proglang{R} \citep{R} packages available from the Comprehensive R
Archive Network (CRAN) aimed solely at simulating survival data, namely
the \pkg{survsim} \citep{survsim}, \pkg{simsurv} \citep{simsurv} and the
\pkg{PermAlgo} \citep{PermAlgo} packages.

The package \pkg{survsim} permits the generation of survival times from
accelerated failure time models, as well as survival data with multiple
events and recurrent events. However, only a restricted set of baseline
distributions that includes the Weibull, log-normal, and log-logistic
distributions is available in this package. The package \pkg{simsurv},
on the other hand, allows the simulation of survival times from
exponential, Weibull, and Gompertz baseline distributions. It further
permits simulating from user-defined hazard functions, and from
two-component mixture distributions. In the package \pkg{simsurv},
covariates are introduced under a proportional hazards assumption, and
time-dependent covariates can also be included by interacting covariates
with linear time or some user-defined transformation of time. Finally,
the package \pkg{PermAlgo} uses a permutational algorithm for the
generation of survival data conditional on an user-specified list of
possibly time-dependent covariates. Existing routines, implemented in
other packages available on CRAN to simulating survival data under more
restrictive settings include the \pkg{mlt} \citep{mlt}, \pkg{prodlim}
\citep{prodlim}, and the \pkg{SimHaz} \citep{SimHaz} packages.

A common feature shared by the packages discussed above it that they are
only suitable to simulating right-censored survival data under an
independent censoring mechanism. In addition, the data generation
functions implemented in those packages return a data.frame as output.
Those facts, unfortunately, forces the simulated data set to fit in some
specific characteristics.

In this paper we introduce the \proglang{R} package \pkg{rsurv}
\citep{rsurv}, available from CRAN at
\url{http://CRAN.R-project.org/package=rsruv}. The package \pkg{rsurv}
is aimed to simulating survival data from some of the main parametric
survival regression models available in the literature, under a variate
of censoring scenarios. The main advantages of the \proglang{R} package
\pkg{rsurv} over other existing packages are:

\begin{enumerate}
\def\labelenumi{\arabic{enumi})}
\item
  The explicit use of R formulas for the specification of the linear
  predictor, facilitating the inclusion of interaction terms and offset
  variables.
\item
  Its versatility to simulating from virtually any baseline survival
  distribution, provided that it possesses a quantile function
  implemented in R.
\item
  The implementation of a wide range of regression models, including
  accelerated failure time (AFT), proportional hazards (PH),
  proportional odds (PO), accelerated hazard (AH), Yang and Prentice
  (YP), and extended hazard (EH) models.
\item
  The possibility to generate left, right and interval-censored survival
  data, under the assumption of both independent and dependent censoring
  mechanisms.
\end{enumerate}

The main data generation functions available in the package \pkg{rsurv}
are implemented to return a vector of failure times as output. This,
combined with the use of the \pkg{dplyr} verbs, brings much more
flexibility in the survival data simulation process. As it shall be
demonstrated in this paper, simulating survival data with more complex
structures such as clustered survival data, multivariate survival data,
or survival data with a cure fraction, can be easily performed using
those core functions. The only drawback of the package \pkg{rsurv}
regards its inability to simulate survival data in the presence of
time-dependent covariates. This fact, however, seems to be a fair
compromise considering the wide range baseline distributions and
regression models that can be combined to simulate survival data using
the package \pkg{rsurv}.

The paper is organized as follows. In Section \ref{sec-core}, we present
the main algorithm implemented in the package \pkg{rsurv}, along with
the core functions used for survival data generation. In Section
\ref{sec-complex}, we illustrate how the package \pkg{rsurv} can be used
to simulate survival data with more complex structures, including
survival data with cure fraction, survival data with random effects
(frailties), multivarite survival data, and competing risks survival
data. Finally, in Section \ref{sec-conclusions} we draw some concluding
remarks and discuss some possible extensions for the proposed package.

\section{Simulation from parametric survival regression
models}\label{sec-core}

Let \(S(t|\boldsymbol{\Theta}, \mathbf{x})\) be the survival function
associated with a given parametric survival regression model, where
\(\boldsymbol{\Theta}\) is a vector of parameters and \(\mathbf{x}\) is
a \(1 \times p\) vector of covariates. Since
\(S(t|\boldsymbol{\Theta}, \mathbf{x}) \sim U(0,1)\), it follows that
samples from a regression model with survival function
\(S(t|\boldsymbol{\Theta}, \mathbf{x})\) can be generated considering
the following equation:

\begin{eqnarray}
\label{TinvS}
T = S^{-1}(U|\boldsymbol{\Theta}, \mathbf{x}),
\end{eqnarray} where \(U \sim U(0,1)\).

Equation \eqref{TinvS} forms the basis for any simulation procedure
aimed for simulating from standard parametric survival regression
models. It is important to note that, for data generation purposes, the
inverse of \(S(t|\boldsymbol{\Theta}, \mathbf{x})\) does not need to
have a closed-form expression provided that it can be evaluated
numerically.

Any regression model can be thought of as a mechanism by which the
distribution associated with a homogeneous population is perturbed in
some fashion by the information contained in a set of covariates. In
survival analysis, such distribution is commonly referred to as the
baseline distribution. Thus, if \(S_{0}(\cdot|\boldsymbol{\theta})\)
corresponds to a baseline survival function depending on a vector of
parameters \(\boldsymbol{\theta}\), then it is reasonable to expect that
\(S(t|\boldsymbol{\Theta}, \mathbf{x}) = S_{0}(t|\boldsymbol{\theta})\)
when \(\mathbf{x} = \mathbf{0}\).

Under appropriate constraints on its regression coefficients, it is
possible to show that the YP model includes the PH and PO models as
particular cases. Similarly, the AFT, AH, and PH models arise as special
cases of the EH model. Therefore, to simulate from all those models it
is sufficient to know how to simulate from the YP and EH models. In the
next sections, we present the basics of the YP and EH models needed for
simulation purposes.

\subsection{YP model and its particular
cases}\label{yp-model-and-its-particular-cases}

The YP model corresponds to a suitable choice to model survival data
with crossing survival curves. This model may also be appropriated in
situations where PH and PO assumptions are not satisfied.

Following \cite{2021_Demarqui} and references therein, the survival
function of the Yang and Prentice model is given by:

\begin{equation}
\label{survYP}
S(t|\boldsymbol{\Theta}, \mathbf{x}) = \left[1+R_{0}(t|\boldsymbol{\theta})e^{\mathbf{x}(\boldsymbol{\beta}-\boldsymbol{\phi})}\right]^{-e^{\mathbf{x}\boldsymbol{\phi}}},
\end{equation}

where
\(\displaystyle R_{0}(t|\boldsymbol{\theta}) = \frac{1-S_{0}(t|\boldsymbol{\theta})}{S_{0}(t|\boldsymbol{\theta})}\)
is the baseline odds function, \(\boldsymbol{\beta}\) and
\(\boldsymbol{\phi}\) are \(p \times 1\) vectors of regression
coefficients.

The hazard function of the YP model can be expressed as:

\begin{equation}
\nonumber
\label{htYP}
  h(t|\boldsymbol{\Theta}, \mathbf{x}) =  \frac{e^{\mathbf{x}(\boldsymbol{\beta} + \boldsymbol{\phi})}}{e^{\mathbf{x}\boldsymbol{\beta}} F_{0}(t|\boldsymbol{\theta})+e^{\mathbf{x}\boldsymbol{\phi}} S_{0}(t|\boldsymbol{\theta})}h_{0}(t|\boldsymbol{\theta}).
\end{equation}

Let \(\mathbf{x}_{1}\) and \(\mathbf{x}_{2}\) two vectors of covariates.
Becase

\begin{eqnarray}
\nonumber
  \lim_{t \rightarrow 0} \frac{h(t|\boldsymbol{\Theta}, \mathbf{x}_{2})}{h(t|\boldsymbol{\Theta}, \mathbf{x}_{1})} = \exp\{(\mathbf{x}_{2} - \mathbf{x}_{1})\boldsymbol{\beta}\}
\end{eqnarray} and \begin{eqnarray}
\nonumber
  \lim_{t \rightarrow \infty} \frac{h(t|\boldsymbol{\Theta}, \mathbf{x}_{2})}{h(t|\boldsymbol{\Theta}, \mathbf{x}_{1})} = \exp\{(\mathbf{x}_{2} - \mathbf{x}_{1})\boldsymbol{\phi}\},
\end{eqnarray} \(\boldsymbol{\beta}\) and \(\boldsymbol{\phi}\) are
often called short and log term regression coefficients, respectively.

It can be show that the survival curves cross each other when
\(\beta_{j} \times \phi_{j} < 0\), for some \(j=1, \cdots, p\).
Moreover, the YP model includes the PH and PO models as particular cases
when \(\boldsymbol{\beta} = \boldsymbol{\phi}\) and
\(\boldsymbol{\phi} = \mathbf{0}\), respectively. For this reason, the
inverse of the survival function given in \eqref{survYP} allows the
simulation from the PH, PO and YP models.

\subsection{EH model and its particular
cases}\label{eh-model-and-its-particular-cases}

The EH model, originally proposed by \cite{1987_Amoli}\ldots{}

\begin{equation}
\nonumber
\label{hazEHorig}
h(t|\boldsymbol{\Theta}, \mathbf{x}) = h_{0}\left(te^{\mathbf{x}\boldsymbol{\beta}}|\boldsymbol{\theta}\right)e^{\mathbf{x}\boldsymbol{\phi}}.
\end{equation}

According to \cite{2011_Tseng}, the regression coefficients
\(\boldsymbol{\beta}\) can be regarded as acceleration factors since
they identify whether the acceleration or deceleration of the observed
failure time related to the baseline failure time. In contrast, the
regression coefficients \(\boldsymbol{\phi}\) characterize the relative
hazard after transforming the time scale of the baseline hazard into an
observed time scale.

In this paper we consider a different parametrization for the EH, in
line with the parametrization adopted in the \proglang{R} package
\pkg{survstan} \citep{survstan}. Under that parametrization, the hazard
and survival functions of the EH model can be expressed as:

\begin{equation}
\label{hazEH}
\nonumber
h(t|\boldsymbol{\Theta}, \mathbf{x}) = h_{0}\left(t/e^{\mathbf{x}\boldsymbol{\beta}}|\boldsymbol{\theta}\right)e^{\mathbf{x}\boldsymbol{\phi}}
\end{equation}

and

\begin{eqnarray}
\label{survEH}
S(t|\boldsymbol{\Theta}, \mathbf{x}) &=& \exp\left\{-H_{0}\left(t/e^{\mathbf{x}\boldsymbol{\beta}}|\boldsymbol{\theta})e^{\mathbf{x}(\boldsymbol{\beta} + \boldsymbol{\phi}}\right)\right\} \nonumber \\
&=& S_{0}\left(t/e^{\mathbf{x}\boldsymbol{\beta}}|\boldsymbol{\theta}\right)^{e^{\mathbf{x}(\boldsymbol{\beta} + \boldsymbol{\phi})}}.
\end{eqnarray}

It is easy to see from Equation \eqref{survEH} that the EH models
includes the AFT, PH and AH models as particular cases when
\(\boldsymbol{\phi} = -\boldsymbol{\beta}\),
\(\boldsymbol{\beta} = \mathbf{0}\), and
\(\boldsymbol{\phi} = \mathbf{0}\), respectively. Therefore, the inverse
of the survival function given in \eqref{survEH} allows the simulation
from the AFT, AH, PH and EH models.

\subsection{Survival data generation procedure}\label{sec-algorithm}

The discussion presented in the previous sections suggests that the
survival function \(S(t|\boldsymbol{\Theta}, \mathbf{x})\) should be
expressed as some function of \(S_{0}(t|\boldsymbol{\theta})\) and
\(\mathbf{x}\), satisfying
\(S(t|\boldsymbol{\Theta}, \mathbf{x}) = S_{0}(t|\boldsymbol{\theta})\)
when \(\mathbf{x} = \mathbf{0}\). The following theorem formalize this
idea, and establishes how random samples from the YP and EH regression
models (and its particular cases AFT, AH, PH and PO) can be draw.

\textbf{Theorem 1} Let \(\boldsymbol{\eta} = (\eta_{1}, \eta_{2})\),
where \(\eta_{1} = \mathbf{x}\boldsymbol{\beta}\) and
\(\eta_{2} = \mathbf{x}\boldsymbol{\phi}\), and consider the survival
functions given in Equations \eqref{survYP} and \eqref{survEH}. Then:

\begin{itemize}
  \item[i)] There exist functions $g(s) = g(s; \boldsymbol{\eta})$ and $a(t) = a(t;\boldsymbol{\eta})$, satisfying $S(t|\boldsymbol{\Theta}, \mathbf{x}) = S_{0}(t|\boldsymbol{\theta})$ when $\mathbf{x} = \mathbf{0}$, such that:

\begin{equation}
\label{surv}
  S(t|\boldsymbol{\Theta}, \mathbf{x}) = g\left(S_{0}(a(t)|\boldsymbol{\theta}\right)).
\end{equation}

  \item[ii)] Samples of $T \sim S(t|\boldsymbol{\Theta}, \mathbf{x})$ can be generated by:
\begin{equation}
\label{simT}
  T = a^{-1}\left(S_{0}^{-1}\left(g^{-1}(U)|\boldsymbol{\theta}\right)\right),
\end{equation}
where $U \sim U(0,1)$.

\end{itemize}

The proof of the theorm is given in the appendix.

In practice, the validity of Equation \eqref{simT} for simulating
survival data from the regression models presented in the previous
section is ensured by two mild general assumptions:

\begin{itemize}
  \item[i)] There is available an R function that returns the inverse of $S_{0}(\cdot|\boldsymbol{\theta})$, that is,  $t = S_{0}^{-1}(u|\boldsymbol{\theta})$, for $u \in (0, 1)$.
  \item[ii)] The linear predictors $\eta_{1} = \mathbf{x}\boldsymbol{\beta}$ and  $\eta_{2} = \mathbf{x}\boldsymbol{\phi}$ do not include a intercept term.
\end{itemize}

It is important to notice that assumption i) does not require the
existence of a closed-form expression for
\(S_{0}(\cdot|\boldsymbol{\theta})\), as long as its inverse can be
computed numerically. Assumption ii), on the other hand, is necessary to
avoid identifiability issues since no further assumptions/constraints
are made about the baseline survival function
\(S_{0}(\cdot|\boldsymbol{\theta})\).

As one can see, the result given in Equation \eqref{simT} is quite
general, and can be applied to simulate data from a wide range of
survival models. The only restriction for its use regards the fact that
neither covariates nor their effects are allowed to vary with time.

The internal function \code{qsurv}, reproduced below, plays a distinct
role in the algorithms implemented in the package \pkg{rsurv}.
Specifically, this function allows the use of any baseline survival
distribution for which there exists the implementation of its quantile
function.

\begin{Shaded}
\begin{Highlighting}[]
\NormalTok{qsurv }\OtherTok{\textless{}{-}} \ControlFlowTok{function}\NormalTok{(p, baseline, ...)\{}
\NormalTok{  qfunc }\OtherTok{\textless{}{-}} \FunctionTok{get}\NormalTok{(}\FunctionTok{paste}\NormalTok{(}\StringTok{"q"}\NormalTok{, baseline, }\AttributeTok{sep =} \StringTok{""}\NormalTok{), }\AttributeTok{mode =} \StringTok{"function"}\NormalTok{)}
\NormalTok{  x }\OtherTok{\textless{}{-}} \FunctionTok{qfunc}\NormalTok{(p, }\AttributeTok{lower.tail =} \ConstantTok{FALSE}\NormalTok{, ...)}
  \FunctionTok{return}\NormalTok{(x)}
\NormalTok{\}}
\end{Highlighting}
\end{Shaded}

The following example shows the usefulness of the \code{qsurv} function
when working with customized functions or functions imported from other
packages. There, the function \code{qmydist} mimics the function
\code{qexp} available in the package \pkg{stats}, whereas the function
\code{qgengamma.orig} from the package \pkg{flexsurv} returns the
quantile function of the generalized gamma distribution, which includes
the exponential distribution as a particular case when all parameters
are equal to 1.

\begin{Shaded}
\begin{Highlighting}[]
\FunctionTok{library}\NormalTok{(flexsurv)}

\NormalTok{qmydist }\OtherTok{\textless{}{-}} \ControlFlowTok{function}\NormalTok{(p, lambda, ...)\{}
\NormalTok{  x }\OtherTok{\textless{}{-}} \FunctionTok{qexp}\NormalTok{(p, }\AttributeTok{rate =}\NormalTok{ lambda, ...)}
  \FunctionTok{return}\NormalTok{(x)}
\NormalTok{\}}

\FunctionTok{set.seed}\NormalTok{(}\DecValTok{1234567890}\NormalTok{)}
\NormalTok{u }\OtherTok{\textless{}{-}} \FunctionTok{runif}\NormalTok{(}\DecValTok{5}\NormalTok{)}
\NormalTok{x1 }\OtherTok{\textless{}{-}} \FunctionTok{qexp}\NormalTok{(u, }\AttributeTok{rate =} \DecValTok{1}\NormalTok{, }\AttributeTok{lower.tail =} \ConstantTok{FALSE}\NormalTok{)}
\NormalTok{x2 }\OtherTok{\textless{}{-}}\NormalTok{ rsurv}\SpecialCharTok{:::}\FunctionTok{qsurv}\NormalTok{(u, }\AttributeTok{baseline =} \StringTok{"exp"}\NormalTok{, }\AttributeTok{rate =} \DecValTok{1}\NormalTok{)}
\NormalTok{x3 }\OtherTok{\textless{}{-}}\NormalTok{ rsurv}\SpecialCharTok{:::}\FunctionTok{qsurv}\NormalTok{(u, }\AttributeTok{baseline =} \StringTok{"mydist"}\NormalTok{, }\AttributeTok{lambda =} \DecValTok{1}\NormalTok{)}
\NormalTok{x4 }\OtherTok{\textless{}{-}}\NormalTok{ rsurv}\SpecialCharTok{:::}\FunctionTok{qsurv}\NormalTok{(u, }\AttributeTok{baseline =} \StringTok{"gengamma.orig"}\NormalTok{, }\AttributeTok{shape=}\DecValTok{1}\NormalTok{, }\AttributeTok{scale=}\DecValTok{1}\NormalTok{, }\AttributeTok{k=}\DecValTok{1}\NormalTok{)}
\FunctionTok{cbind}\NormalTok{(x1, x2, x3, x4)}
\end{Highlighting}
\end{Shaded}

\begin{verbatim}
##              x1         x2         x3         x4
## [1,] 0.09339179 0.09339179 0.09339179 0.09339179
## [2,] 0.95202641 0.95202641 0.95202641 0.95202641
## [3,] 0.17411789 0.17411789 0.17411789 0.17411789
## [4,] 0.29132206 0.29132206 0.29132206 0.29132206
## [5,] 0.34595405 0.34595405 0.34595405 0.34595405
\end{verbatim}

The regression models implemented in the package \pkg{rsurv} share the
same parametrizations of those implement in the \proglang{R} package
\pkg{survstan} \citep{survstan}. The main functions available in the
package \pkg{rsurv} are: \code{raftreg()}, \code{rahtreg()},
\code{rapheg()}, \code{raporeg()}, \code{rehreg()} and \code{rypreg()}.
These functions permit the simulation of survival times from the AFT,
AH, PH, PO, EH and YP models, respectively.

All \code{r*reg()} functions share the same basic structure.
Specifically, the user must provide a numeric vector \code{u} of
quantiles used in \eqref{simT}, a \code{formula} containing information
regarding the structure of the linear predictors, a vector \code{beta}
of regression coefficients (and additionally a vector \code{phi} for the
EH and YP models), which should be compatible with the model matrix
induced by the \code{formula} argument, a \code{dist/baseline} argument
used to specify the chosen baseline distribution, along with any
additional arguments passed to the internal function \code{qsurv}
through the \code{...} argument.

A distinct feature of the functions \code{r*reg()} is that they return a
vector of generated survival times rather than a full data set
containing additional information regarding covariates and a failure
indicator variable. This brings more flexibility for the user to
simulate survival data with different characteristics and levels of
complexity, as we shall demonstrate throughout the examples presented in
the paper.

As a first example, suppose we want to simulate a sample of type I
right-censored survival data, assuming that the failure times are
generated from an accelerated failure time model with loglogistic
baseline distribution. In addition, assume that we wish to consider two
exploratory variables, say age and sex, and we want to include an
interaction effect between them. Such a task can be easily accomplished
by using the function \code{raftreg} along with the function
\code{qllogis} available in the package \pkg{flexsurv}, as illustrated
below:

\begin{Shaded}
\begin{Highlighting}[]
\FunctionTok{library}\NormalTok{(flexsurv)}
\FunctionTok{library}\NormalTok{(rsurv)}
\FunctionTok{library}\NormalTok{(survstan)}

\FunctionTok{set.seed}\NormalTok{(}\DecValTok{1234567890}\NormalTok{)}

\NormalTok{n }\OtherTok{\textless{}{-}}  \DecValTok{1000}
\NormalTok{tau }\OtherTok{\textless{}{-}} \DecValTok{10}  \CommentTok{\# maximum follow up time}
\NormalTok{simdata }\OtherTok{\textless{}{-}} \FunctionTok{data.frame}\NormalTok{(}
  \AttributeTok{age =} \FunctionTok{rnorm}\NormalTok{(n),}
  \AttributeTok{sex =} \FunctionTok{sample}\NormalTok{(}\FunctionTok{c}\NormalTok{(}\StringTok{"f"}\NormalTok{, }\StringTok{"m"}\NormalTok{), }\AttributeTok{size =}\NormalTok{ n, }\AttributeTok{replace =} \ConstantTok{TRUE}\NormalTok{)}
\NormalTok{) }\SpecialCharTok{|\textgreater{}}
  \FunctionTok{mutate}\NormalTok{(}
    \AttributeTok{t =} \FunctionTok{raftreg}\NormalTok{(}\FunctionTok{runif}\NormalTok{(n), }\SpecialCharTok{\textasciitilde{}}\NormalTok{ age}\SpecialCharTok{*}\NormalTok{sex, }\AttributeTok{beta =} \FunctionTok{c}\NormalTok{(}\DecValTok{1}\NormalTok{, }\DecValTok{2}\NormalTok{, }\SpecialCharTok{{-}}\FloatTok{0.5}\NormalTok{), }
                \AttributeTok{dist =} \StringTok{"llogis"}\NormalTok{, }\AttributeTok{shape =} \FloatTok{1.5}\NormalTok{, }\AttributeTok{scale =} \DecValTok{1}\NormalTok{),}
\NormalTok{  ) }\SpecialCharTok{|\textgreater{}}
  \FunctionTok{rowwise}\NormalTok{() }\SpecialCharTok{|\textgreater{}}
  \FunctionTok{mutate}\NormalTok{(}
    \AttributeTok{time =} \FunctionTok{min}\NormalTok{(t, tau),}
    \AttributeTok{status =} \FunctionTok{as.numeric}\NormalTok{(time }\SpecialCharTok{==}\NormalTok{ t)}
\NormalTok{  ) }

\FunctionTok{glimpse}\NormalTok{(simdata)}
\end{Highlighting}
\end{Shaded}

\begin{verbatim}
## Rows: 1,000
## Columns: 5
## Rowwise: 
## $ age    <dbl> 1.34592454, 0.99527131, 0.54622688, -1.91272392, 1.92128431, 1.~
## $ sex    <chr> "m", "f", "f", "m", "m", "m", "m", "f", "f", "m", "f", "m", "m"~
## $ t      <dbl> 15.2363453, 1.5259533, 2.1783746, 2.4354995, 58.7932958, 16.714~
## $ time   <dbl> 10.0000000, 1.5259533, 2.1783746, 2.4354995, 10.0000000, 10.000~
## $ status <dbl> 0, 1, 1, 1, 0, 0, 1, 1, 1, 1, 1, 1, 1, 1, 1, 1, 1, 1, 0, 1, 1, ~
\end{verbatim}

\begin{Shaded}
\begin{Highlighting}[]
\NormalTok{fit }\OtherTok{\textless{}{-}} \FunctionTok{aftreg}\NormalTok{(}
  \FunctionTok{Surv}\NormalTok{(time, status) }\SpecialCharTok{\textasciitilde{}}\NormalTok{ age}\SpecialCharTok{*}\NormalTok{sex,}
  \AttributeTok{data =}\NormalTok{ simdata, }\AttributeTok{dist =} \StringTok{"loglogistic"}
\NormalTok{)}
\FunctionTok{estimates}\NormalTok{(fit)}
\end{Highlighting}
\end{Shaded}

\begin{verbatim}
##        age       sexm   age:sexm      alpha      gamma 
##  0.9494630  2.0094422 -0.4812641  1.4946497  1.0226847
\end{verbatim}

In our next example we demonstrate how to generate survival data with
crossing survival curves using the YP model. In this example, we assume
that the survival times are randomly right-censored, with the censoring
times following an exponential distribution.

\begin{Shaded}
\begin{Highlighting}[]
\FunctionTok{library}\NormalTok{(GGally)}
\FunctionTok{set.seed}\NormalTok{(}\DecValTok{1234567890}\NormalTok{)}
\NormalTok{n }\OtherTok{\textless{}{-}} \DecValTok{1000}
\NormalTok{simdata }\OtherTok{\textless{}{-}} \FunctionTok{data.frame}\NormalTok{(}
  \AttributeTok{trt =} \FunctionTok{sample}\NormalTok{(}\FunctionTok{c}\NormalTok{(}\StringTok{"chemo"}\NormalTok{, }\StringTok{"chemo+rad"}\NormalTok{), }\AttributeTok{size =}\NormalTok{ n, }\AttributeTok{replace =} \ConstantTok{TRUE}\NormalTok{)}
\NormalTok{) }\SpecialCharTok{|\textgreater{}}
  \FunctionTok{mutate}\NormalTok{(}
    \AttributeTok{t =} \FunctionTok{rypreg}\NormalTok{(}\FunctionTok{runif}\NormalTok{(n), }\SpecialCharTok{\textasciitilde{}}\NormalTok{ trt, }\AttributeTok{beta =} \DecValTok{2}\NormalTok{, }\AttributeTok{phi =} \SpecialCharTok{{-}}\FloatTok{1.5}\NormalTok{, }
               \AttributeTok{dist =} \StringTok{"weibull"}\NormalTok{, }\AttributeTok{shape =} \FloatTok{1.5}\NormalTok{, }\AttributeTok{scale =} \DecValTok{1}\NormalTok{),}
    \AttributeTok{c =} \FunctionTok{rexp}\NormalTok{(n, }\AttributeTok{rate =} \DecValTok{1}\NormalTok{)}
\NormalTok{  ) }\SpecialCharTok{|\textgreater{}}  
  \FunctionTok{rowwise}\NormalTok{() }\SpecialCharTok{|\textgreater{}}
  \FunctionTok{mutate}\NormalTok{(}
    \AttributeTok{time =} \FunctionTok{min}\NormalTok{(t, c),}
    \AttributeTok{status =} \FunctionTok{as.numeric}\NormalTok{(time }\SpecialCharTok{==}\NormalTok{ t)}
\NormalTok{  ) }\SpecialCharTok{|\textgreater{}} \FunctionTok{select}\NormalTok{(}\SpecialCharTok{{-}}\FunctionTok{c}\NormalTok{(t, c))}

\FunctionTok{glimpse}\NormalTok{(simdata)}
\end{Highlighting}
\end{Shaded}

\begin{verbatim}
## Rows: 1,000
## Columns: 3
## Rowwise: 
## $ trt    <chr> "chemo", "chemo", "chemo+rad", "chemo+rad", "chemo+rad", "chemo~
## $ time   <dbl> 0.65743632, 1.14638933, 0.10715893, 0.09876511, 1.44704010, 0.1~
## $ status <dbl> 1, 0, 1, 1, 1, 0, 1, 0, 1, 1, 0, 1, 0, 0, 1, 1, 0, 1, 0, 1, 0, ~
\end{verbatim}

\begin{Shaded}
\begin{Highlighting}[]
\NormalTok{fit }\OtherTok{\textless{}{-}} \FunctionTok{ypreg}\NormalTok{(}
  \FunctionTok{Surv}\NormalTok{(time, status) }\SpecialCharTok{\textasciitilde{}}\NormalTok{ trt,}
  \AttributeTok{data =}\NormalTok{ simdata, }\AttributeTok{dist =} \StringTok{"weibull"}
\NormalTok{)}
\FunctionTok{estimates}\NormalTok{(fit)}
\end{Highlighting}
\end{Shaded}

\begin{verbatim}
## short-trtchemo+rad  long-trtchemo+rad              alpha              gamma 
##           1.920795          -1.446502           1.404515           1.043123
\end{verbatim}

\begin{Shaded}
\begin{Highlighting}[]
\NormalTok{km }\OtherTok{\textless{}{-}} \FunctionTok{survfit}\NormalTok{(}\FunctionTok{Surv}\NormalTok{(time, status) }\SpecialCharTok{\textasciitilde{}}\NormalTok{ trt, }\AttributeTok{data =}\NormalTok{ simdata)}
\FunctionTok{ggsurv}\NormalTok{(km) }\SpecialCharTok{+} 
  \FunctionTok{theme}\NormalTok{(}\AttributeTok{legend.position=}\StringTok{"bottom"}\NormalTok{)}
\end{Highlighting}
\end{Shaded}

\includegraphics{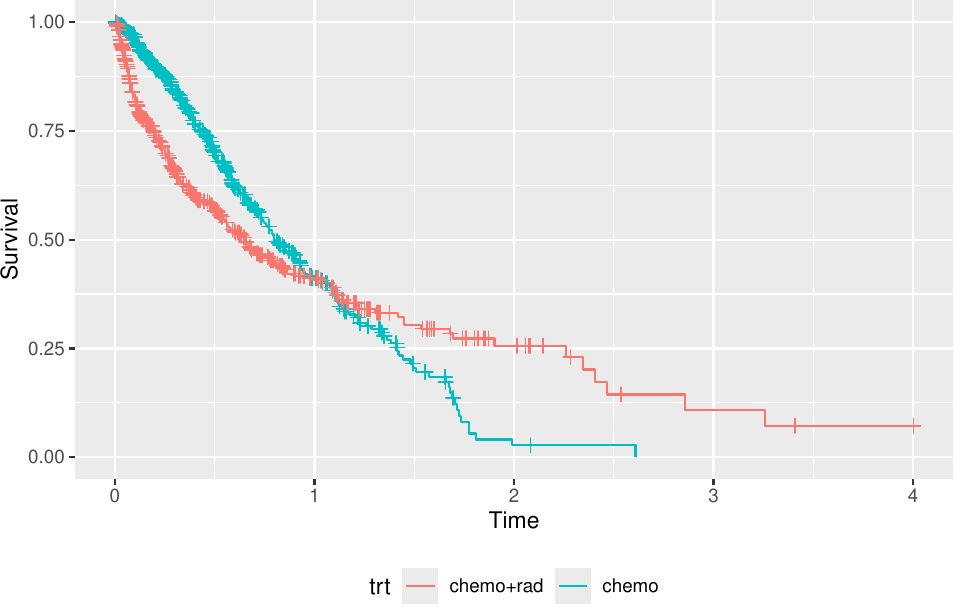}

The two examples presented above clearly demonstrate the flexibility
provided by the functions \code{r*reg()} in the simulation of survival
data with different types of censoring. In the next section we look at
the simulation of survival data with more sophisticated structures.

\section{Simulation of survival data with more complex
structures}\label{sec-complex}

This section is dedicated to show how the functions discussed in the
previous section can be used to simulate survival data with more complex
structures. It is worth noting that the examples presented here only
cover a subset of the simulation possibilities.

The package \pkg{rsurv} contains some specialized functions that are
useful for simulating clustered survival data, survival data with cure
fraction, and interval-censored survival data. Other types of survival
data, such as multivariate survival data, survival data with competing
risks, and survival data with dependent censoring can be generated, for
instance, combining the functions implemented in the package \pkg{rsurv}
with functions available in the \proglang{R} package \pkg{copula}
\citep{copula2, copula}.

\subsection{Cure rate models}\label{cure-rate-models}

\cite{2009Rodrigues} presented an unified formulation for cure rate
models, through the use of probability-generating functions, that
includes several cure rate models available in the literature. In their
framework, cure rate models are formulated based on a two-stage process
that possesses an appealing biological motivation in terms of
incidence-latency of disease.

Suppose we are interested in modeling the time to relapse to cancer and
let \(N\) denote the number of carcinogenic cells (often called
clonogens) left active after a patient has gone through initial
treatment. Then, for an individual with \(N\) clonogens left active, let
\(\xi_{j}\) be the non-observable random time (\emph{i.e.}, the
promotion time) for the \(j\)-th clonogen to produce a detectable cancer
mass, \(j=0,1, ..., N\).

The time to relapse to cancer, which is in fact the observable quantity,
is defined as:

\begin{eqnarray}
T &=&  \infty I\{N=0\} + \xi_{(1)}I\{N  \ge 1\},
\end{eqnarray} where \(\xi_{(1)} <  \xi_{(2)}, < \cdots < \xi_{(N)}\)
denote the orded values of \(\xi_{1}, \xi_{2}, \cdots, \xi_{N}\).

Then, under the assumptions that \(N\) is independent of
\(\xi_{j}, ~ \forall ~j \ge 0\), and
\(\xi_{j} \overset{\text{i.i.d.}}{\sim} S_{\xi}(\cdot|\boldsymbol{\Theta}, \mathbf{x})\),
\(j \ge 0\), it follows that the population survival function can be
expressed as:

\begin{eqnarray}
\label{St_pop}
    S_{pop}(t) &=& P(N=0) + P(\xi_{1}>t, ..., \xi_{N}>t; N>0) \nonumber \\
               &=& P(N=0) + \sum_{k \ge 1}P(N=k)S_{\xi}(t|\boldsymbol{\Theta}, \mathbf{x})^{k}. 
\end{eqnarray}

\textbf{Theorem 2} (\cite{2009Rodrigues}) The survival function
associated with a r.v. \(T\) corresponding to a promotion time
(two-stage) cure rate model is given by: \begin{eqnarray}
  S_{pop}(t) = \sum_{k=0}^{\infty}p_{k}\{S_{\xi}(t|\boldsymbol{\Theta}, \mathbf{x})\}^{k} = A_{p}\left(S_{\xi}(t|\boldsymbol{\Theta}, \mathbf{x})\right),
 \end{eqnarray} where \(p_{k}=P(N=k)\) and \(A_{p}(\cdot)\) is the
probability-generating function of \(N\).

The next corollary of Theorems 1 and 2 provides a general expression to
simulate survival data with a cure fraction.

\textbf{Corollary 1} Given a cure fraction probability \(\pi = P(N=0)\)
and a survival function \(S_{\xi}(t|\boldsymbol{\Theta}, \mathbf{x})\)
satisfying \eqref{surv}, then time to relapse to cancer \(T\) can be
generated as follows:

\begin{equation}
\label{t_cure}
T = \left.
  \begin{cases}
    \infty, & \text{if } U < \pi \\
    a^{-1}\left(g^{-1}\left(A_{p}^{-1}(U)\right)\right), & \text{if } U \ge \pi
  \end{cases},
  \right.
\end{equation}

where \(U \sim U(0, 1)\).

The proof of the corollary in show in the Appendix.

It is important to note from Equation \eqref{t_cure} that, given
\(V = A_{p}^{-1}(U)\), all the time generation procedures described
previously can be straightforwardly applied in the generation of
survival times in the presence of a cure fraction, provided that the
inverse of the probability generating function, \(A_{p}^{-1}(u)\), is
known.

\begin{table}
\label{tbl_inv_pgf}
\begin{center}
  \caption{$A_{p}(s)$ and $A_{p}^{-1}(u)$ expressions.}
  \begin{tabular}{cccc} \hline
  distribution & $A_{p}(s)$ & $A_{p}^{-1}(u)$ \\ \hline
  Bernoulli & $(1-\mu) + \mu s$ & $(u - 1 +\mu)/\mu$ \\
  Poisson   & $\exp\{-\mu(1-s)\}$ & $[\log(\mu) + \mu]/\mu$\\
  Negative Binomial    & $1 - (u^{-\zeta} - 1)/\zeta\mu$ & $\left[1+\zeta\mu\right]^{-1/\zeta}$\\
  Bell      & $\exp\left\{e^{s\theta}-e^{\theta}\right\}$ & $\log(\log(u) +e^{\theta})$\\ \hline 
\end{tabular} \\
\footnotesize{Here, $E(N) = \mu$ and $\theta = W_{0}(\mu)$ is the Lambert W function.}
\end{center}
\end{table}

Common distributions assumed for the incidence sub-model (\emph{i.e.},
the distribution of the latent r.v. \(N\)) are the Bernoulli, Poisson,
negative binomial and the Bell distributions. Covariates are usually
included in the incidence sub-model. If \(\mathbf{z}\) is a
\(1\times q\) vector of covariates, and \(\boldsymbol{\kappa}\) is the
corresponding \(q \times 1\) vector of regression coefficients, than
\(\pi = P(N=0)\) can be related to the linear predictor
\(\mathbf{z}\boldsymbol{\kappa}\) through appropriate choices of link
functions.

The inverse of the probability generating functions presented in Table
\ref{tbl_inv_pgf} are implemented in the function \texttt{inv\_pgf()}.
Similarly to the \code{r*reg()} functions, the function
\texttt{inv\_pgf()} requires a \code{formula} argument specifying the
structure of the linear predictor, along with a \code{kappa} argument
passing the vector of regression coefficients, that should be compatible
with the model matrix induced by the \code{formula} argument, and an
\code{incidence} argument determining the incidence sub-model, and its
corresponding link function. Specifically, the function
\code{rsurv::bernoulli()} allows the specification of logit, probit,
cloglog, and cauchit link functions, whereas the log, identity and
square root links can be used with the functions
\code{rsurv::poisson()}, \code{rsurv::bell()} and
\code{rsurv::negbin()}. When the negative binomial distribution is
assumed for the incidence sub-model, the function \texttt{inv\_pgf()}
further requires an additional argument \code{zeta}, representing the
extra negative-binomial parameter.

The next example illustrates how survival data with cure fraction can be
easily generated using the package \pkg{rsurv}. In that example we
consider a regression structure only for the incidence sub-model. More
specifcally, we simulate survival data from a mixture cure rate model
with a probit link function. We further assume that the survival times
are randomly right-censored.

\begin{Shaded}
\begin{Highlighting}[]
\FunctionTok{library}\NormalTok{(rsurv)}
\FunctionTok{library}\NormalTok{(GGally)}

\CommentTok{\# fixing the seed for the rng:}
\FunctionTok{set.seed}\NormalTok{(}\DecValTok{1234567890}\NormalTok{)}

\NormalTok{kappa }\OtherTok{\textless{}{-}} \FunctionTok{c}\NormalTok{(}\FloatTok{0.5}\NormalTok{, }\FloatTok{1.5}\NormalTok{, }\SpecialCharTok{{-}}\FloatTok{1.1}\NormalTok{)}
\NormalTok{n }\OtherTok{\textless{}{-}} \DecValTok{1000}

\CommentTok{\# generating the set of explanatory variables:}
\NormalTok{simdata }\OtherTok{\textless{}{-}} \FunctionTok{data.frame}\NormalTok{(}
  \AttributeTok{trt =} \FunctionTok{sample}\NormalTok{(}\FunctionTok{c}\NormalTok{(}\StringTok{"A"}\NormalTok{, }\StringTok{"B"}\NormalTok{), }\AttributeTok{size =}\NormalTok{ n, }\AttributeTok{replace =} \ConstantTok{TRUE}\NormalTok{),}
  \AttributeTok{age =} \FunctionTok{rnorm}\NormalTok{(n)}
\NormalTok{)}

\CommentTok{\# generating the data set:}
\NormalTok{v }\OtherTok{\textless{}{-}} \FunctionTok{inv\_pgf}\NormalTok{(}
  \SpecialCharTok{\textasciitilde{}}\NormalTok{ trt }\SpecialCharTok{+}\NormalTok{ age,}
  \AttributeTok{incidence =} \FunctionTok{bernoulli}\NormalTok{(}\StringTok{"probit"}\NormalTok{),}
  \AttributeTok{kappa =}\NormalTok{ kappa,}
  \AttributeTok{data =}\NormalTok{ simdata}
\NormalTok{)}

\NormalTok{simdata }\OtherTok{\textless{}{-}}\NormalTok{ simdata }\SpecialCharTok{|\textgreater{}}
  \FunctionTok{mutate}\NormalTok{(}
    \AttributeTok{t =} \FunctionTok{qexp}\NormalTok{(v, }\AttributeTok{rate =} \DecValTok{1}\NormalTok{, }\AttributeTok{lower.tail =} \ConstantTok{FALSE}\NormalTok{),}
    \AttributeTok{c =} \FunctionTok{rexp}\NormalTok{(n, }\AttributeTok{rate =} \DecValTok{1}\NormalTok{)}
\NormalTok{  ) }\SpecialCharTok{|\textgreater{}}
  \FunctionTok{rowwise}\NormalTok{() }\SpecialCharTok{|\textgreater{}}
  \FunctionTok{mutate}\NormalTok{(}
    \AttributeTok{time =} \FunctionTok{min}\NormalTok{(t, c),}
    \AttributeTok{status =} \FunctionTok{as.numeric}\NormalTok{(time }\SpecialCharTok{==}\NormalTok{ t)}
\NormalTok{  ) }\SpecialCharTok{|\textgreater{}}
  \FunctionTok{select}\NormalTok{(}\SpecialCharTok{{-}}\FunctionTok{c}\NormalTok{(t, c))}

\FunctionTok{glimpse}\NormalTok{(simdata)}
\end{Highlighting}
\end{Shaded}

\begin{verbatim}
## Rows: 1,000
## Columns: 4
## Rowwise: 
## $ trt    <chr> "A", "A", "B", "B", "B", "A", "B", "A", "B", "A", "A", "A", "A"~
## $ age    <dbl> 0.2193297, 0.9952571, -0.4572095, 0.4864704, 1.9512081, 0.49597~
## $ time   <dbl> 0.17789062, 0.17142535, 0.50301782, 0.64774117, 0.40046281, 0.4~
## $ status <dbl> 0, 0, 0, 1, 0, 0, 1, 1, 1, 1, 1, 1, 0, 0, 0, 1, 1, 0, 0, 1, 1, ~
\end{verbatim}

\begin{Shaded}
\begin{Highlighting}[]
\NormalTok{km }\OtherTok{\textless{}{-}} \FunctionTok{survfit}\NormalTok{(}\FunctionTok{Surv}\NormalTok{(time, status) }\SpecialCharTok{\textasciitilde{}}\NormalTok{ trt, }\AttributeTok{data =}\NormalTok{ simdata)}
\FunctionTok{ggsurv}\NormalTok{(km) }\SpecialCharTok{+}
  \FunctionTok{ylim}\NormalTok{(}\DecValTok{0}\NormalTok{, }\DecValTok{1}\NormalTok{)}
\end{Highlighting}
\end{Shaded}

\includegraphics{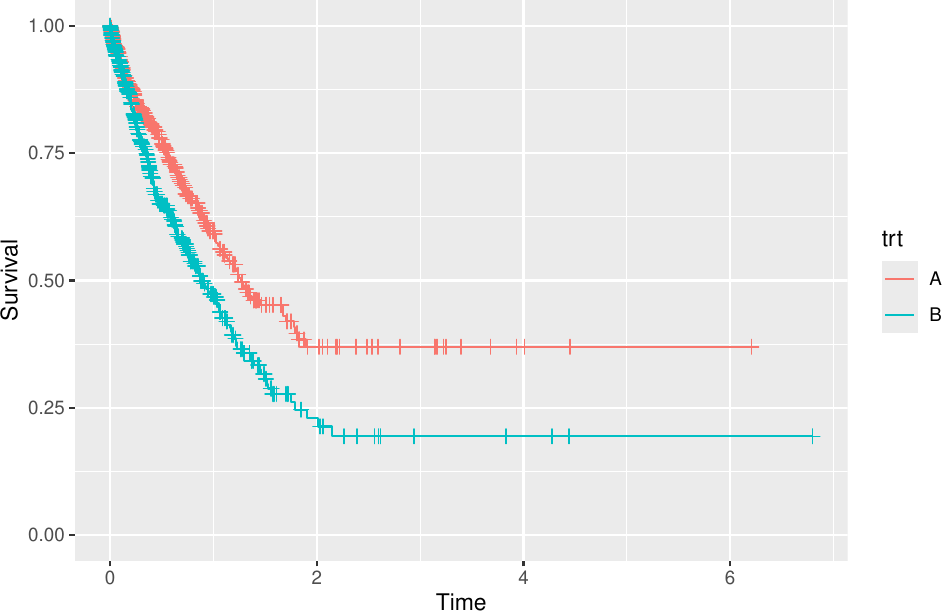}

\subsection{Fraily models}\label{fraily-models}

Frailty models are often employed in survival analysis to accommodate
unknown and/or unobserved risk factors through the inclusion of random
effects, known as frailties (a term coined by \cite{1979_Vaupel} to
represent individual's heterogeneity). These type of models have become
increasingly popular in multivariate survival analysis since they allow
one to take into account the association among individual survival times
within subgroups (or clusters) of subjects.

\begin{table}
\label{tbl_frailties}
\begin{center}
  \caption{Frailty distributions available from the \code{rsurv::frailty()} function.}
  \begin{tabular}{lcc} \hline
  Frailty term & Expected value & Variance \\ \hline
  $w \sim \mbox{N}(0, \sigma^2)$ & 0 & $\sigma^2$ \\
  $z \sim \mbox{Gamma}(1/\sigma^2, 1/\sigma^2)$ & 1 & $\sigma^2$ \\
  $z \sim \mbox{PS}(\alpha)$ & - & - \\ \hline
\end{tabular} \\
\footnotesize{The parameter $0 < \alpha \le 2$ is known as the stability parameter of the PS distribution.}
\end{center}
\end{table}

The package \pkg{rsurv} permits the inclusion of frailty terms through
the linear predictors, \emph{i.e.,}
\(\eta_{1} = \mathbf{x}\boldsymbol{\beta} + w\) and
\(\eta_{2} = \mathbf{x}\boldsymbol{\phi} + w\), where \(w\) is the
frailty term. This convention adopted by the package \pkg{rsruv} allows
the inclusion of frailties into the linear predictors using the
\code{stats::offset()} function.

The function \code{rfrailty()} implemented in the package \pkg{rsurv}
facilitates the process of simulating survival data with frailties. This
function permits random generation of frailties from the most commonly
frailty distributions used in practice, namely the gamma distribution,
the gaussian distribution, and the positive stable (ps) distribution,
which are briefly described in Table \ref{tbl_frailties}. Since the
gamma and positive stable distributions have support on \(\Re^{+}\), by
default, the function \code{rfrailty()} returns samples from positive
random frailties in the log-scale, that is, \(w = \log(z)\).

Our next example shows how the functions \code{r*reg()} and
\code{rfrailty()} can be combined to generate survival data from shared
frailty models. For this type of survival frailty models, elements
belonging to the same cluster share the same random effect (frailty). To
generate frailties using the function \code{rfrailty()} we need to pass
a vector of length \(n\) containing information regarding the cluster
configuration to the \code{cluster} argument, the assumed distribution
to the \code{frailty} argument, and finally, the parameter of the
frailty distribution (\code{sigma} or \code{alpha}, depending on the
choice of passed to \code{frailty} argument).

\begin{Shaded}
\begin{Highlighting}[]
\FunctionTok{library}\NormalTok{(rsurv)}
\FunctionTok{library}\NormalTok{(frailtyEM)}

\FunctionTok{set.seed}\NormalTok{(}\DecValTok{1234567890}\NormalTok{)}

\NormalTok{n }\OtherTok{\textless{}{-}} \DecValTok{1000}  \CommentTok{\# sample size}
\NormalTok{L }\OtherTok{\textless{}{-}} \DecValTok{100}   \CommentTok{\# number of clusters}
\NormalTok{simdata }\OtherTok{\textless{}{-}} \FunctionTok{data.frame}\NormalTok{(}
  \AttributeTok{id =} \FunctionTok{rep}\NormalTok{(}\DecValTok{1}\SpecialCharTok{:}\NormalTok{L, }\AttributeTok{each =}\NormalTok{ n}\SpecialCharTok{/}\NormalTok{L),}
  \AttributeTok{age =} \FunctionTok{rnorm}\NormalTok{(n),}
  \AttributeTok{sex =} \FunctionTok{sample}\NormalTok{(}\FunctionTok{c}\NormalTok{(}\StringTok{"f"}\NormalTok{, }\StringTok{"m"}\NormalTok{), }\AttributeTok{size =}\NormalTok{ n, }\AttributeTok{replace =} \ConstantTok{TRUE}\NormalTok{)}
\NormalTok{) }\SpecialCharTok{|\textgreater{}}
  \FunctionTok{mutate}\NormalTok{(}
    \AttributeTok{frailty =} \FunctionTok{rfrailty}\NormalTok{(}\AttributeTok{cluster =}\NormalTok{ id, }\AttributeTok{frailty =} \StringTok{"gamma"}\NormalTok{, }\AttributeTok{sigma =} \FloatTok{0.5}\NormalTok{),}
    \AttributeTok{t =} \FunctionTok{rphreg}\NormalTok{(}\AttributeTok{u =} \FunctionTok{runif}\NormalTok{(n), }\SpecialCharTok{\textasciitilde{}}\NormalTok{ age}\SpecialCharTok{*}\NormalTok{sex }\SpecialCharTok{+} \FunctionTok{offset}\NormalTok{(frailty) , }
               \AttributeTok{beta =} \FunctionTok{c}\NormalTok{(}\DecValTok{1}\NormalTok{, }\DecValTok{2}\NormalTok{, }\SpecialCharTok{{-}}\FloatTok{0.5}\NormalTok{), }\AttributeTok{dist =} \StringTok{"exp"}\NormalTok{, }\AttributeTok{rate =} \DecValTok{1}\NormalTok{),}
    \AttributeTok{c =} \FunctionTok{runif}\NormalTok{(n, }\DecValTok{0}\NormalTok{, }\DecValTok{5}\NormalTok{)}
\NormalTok{  ) }\SpecialCharTok{|\textgreater{}}
  \FunctionTok{rowwise}\NormalTok{() }\SpecialCharTok{|\textgreater{}}
  \FunctionTok{mutate}\NormalTok{(}
    \AttributeTok{time =} \FunctionTok{min}\NormalTok{(t, c),}
    \AttributeTok{status =} \FunctionTok{as.numeric}\NormalTok{(t}\SpecialCharTok{\textless{}}\NormalTok{c)}
\NormalTok{  ) }\SpecialCharTok{|\textgreater{}}
  \FunctionTok{select}\NormalTok{(}\SpecialCharTok{{-}}\FunctionTok{c}\NormalTok{(t, c))}
\FunctionTok{glimpse}\NormalTok{(simdata)}
\end{Highlighting}
\end{Shaded}

\begin{verbatim}
## Rows: 1,000
## Columns: 6
## Rowwise: 
## $ id      <int> 1, 1, 1, 1, 1, 1, 1, 1, 1, 1, 2, 2, 2, 2, 2, 2, 2, 2, 2, 2, 3,~
## $ age     <dbl> 1.34592454, 0.99527131, 0.54622688, -1.91272392, 1.92128431, 1~
## $ sex     <chr> "m", "f", "f", "m", "m", "m", "m", "f", "f", "m", "f", "m", "m~
## $ frailty <dbl> -0.15917714, -0.15917714, -0.15917714, -0.15917714, -0.1591771~
## $ time    <dbl> 0.066662943, 0.370877454, 3.527734703, 0.109155541, 0.11269257~
## $ status  <dbl> 1, 1, 0, 1, 1, 1, 1, 1, 1, 1, 1, 1, 1, 1, 1, 1, 1, 0, 1, 1, 1,~
\end{verbatim}

\begin{Shaded}
\begin{Highlighting}[]
\NormalTok{em }\OtherTok{\textless{}{-}} \FunctionTok{emfrail}\NormalTok{(}\FunctionTok{Surv}\NormalTok{(time, status) }\SpecialCharTok{\textasciitilde{}}\NormalTok{ age}\SpecialCharTok{*}\NormalTok{sex }\SpecialCharTok{+} \FunctionTok{cluster}\NormalTok{(id),}
              \AttributeTok{distribution =} \FunctionTok{emfrail\_dist}\NormalTok{(}\AttributeTok{dist =} \StringTok{"gamma"}\NormalTok{),}
              \AttributeTok{data =}\NormalTok{ simdata)}
\FunctionTok{summary}\NormalTok{(em)}
\end{Highlighting}
\end{Shaded}

\begin{verbatim}
## Call: 
## emfrail(formula = Surv(time, status) ~ age * sex + cluster(id), 
##     data = simdata, distribution = emfrail_dist(dist = "gamma"))
## 
## Regression coefficients:
##             coef exp(coef) se(coef) adj. se       z p
## age       0.9711    2.6409   0.0675  0.0682 14.2469 0
## sexm      2.1159    8.2973   0.0949  0.0976 21.6777 0
## age:sexm -0.4746    0.6221   0.0811  0.0813 -5.8405 0
## Estimated distribution: gamma / left truncation: FALSE 
## 
## Fit summary:
## Commenges-Andersen test for heterogeneity: p-val  1.25e-19 
## no-frailty Log-likelihood: -4679.899 
## Log-likelihood: -4636.705 
## LRT: 1/2 * pchisq(86.4), p-val 7.39e-21
## 
## Frailty summary:
##                    estimate lower 95% upper 95%
## Var[Z]                0.233     0.152     0.350
## Kendall's tau         0.105     0.071     0.149
## Median concordance    0.102     0.068     0.145
## E[logZ]              -0.121    -0.185    -0.078
## Var[logZ]             0.263     0.164     0.418
## theta                 4.284     2.859     6.570
## Confidence intervals based on the likelihood function
\end{verbatim}

\subsection{Interval-censored survival
data}\label{interval-censored-survival-data}

Consider a nonnegative random variable \(T\) representing the time until
the occorrence of an event of interest. According to \cite{2006_Sun}, an
observation on \(T\) is interval-censored when its exact value is
unknow, and only an interval of the form \((L , R]\) can be observed, so
that

\begin{equation}
\label{eq_icens}
T \in (L, R],
\end{equation} where \(L \le R\).

It is easy to see from \eqref{eq_icens} that right and left-censored
observations on \(T\) occur when \(R = \infty\) and \(L = 0\),
respectively. Naturally, when \(L=R\), \(T\) can be observed exactly.

Two important types of interval-censored survival data that often arise
in practice are the so-called type I (also known as current status
survival data in the literature) and type II interval-censored survival
data. Type I interval-censored survival data occurs when all observed
intervals result in either left-censored (\(L = 0\)) or right-censored
(\(R = \infty\)) observations. Type II interval-censored survival data,
on the other hand, emerges when we have at least one finite interval
satisfying \(L>0\).

The generation of type I and type II interval-censored survival data is
facilitated by the function \code{rinterval} available in the package
\pkg{rsurv}. For the random generation of type II interval-censored
survival data, the algorithm described in \cite{2012_Kiani}, which
emulates a situation in which patients are expected to attend
pre-scheduled visits to the physician with attendance probability \(p\),
is implemented. The function \code{rinterval} requires an argument
\code{time} passing a sample of failure times, an argument \code{tau}
that can be either a vector of censoring times (for type I
interval-censored survival data) or a time-grid of scheduled visits (for
type II interval-censored survival data), and an argument \code{type}
indicating the type of interval-censored survival data. Finally, the
\code{prob} argument, only evaluated when \code{type = II}, can be
regarded as the attendance probability of a scheduled visit.

The code presented below examplifies how to simulate type I
interval-censored survival data using the function function
\code{rinterval()}. There, it is assumed that the exact failure times
are generated from a PH model with an exponential baseline distribution,
whereas the censoring times are generated from a Weibull distribution.

\begin{Shaded}
\begin{Highlighting}[]
\FunctionTok{library}\NormalTok{(rsurv)}
\FunctionTok{library}\NormalTok{(icenReg)}

\FunctionTok{set.seed}\NormalTok{(}\DecValTok{1234567890}\NormalTok{)}
\NormalTok{n }\OtherTok{\textless{}{-}} \DecValTok{300}

\NormalTok{covariates }\OtherTok{\textless{}{-}} \FunctionTok{data.frame}\NormalTok{(}
  \AttributeTok{age =} \FunctionTok{rnorm}\NormalTok{(n),}
  \AttributeTok{sex =} \FunctionTok{sample}\NormalTok{(}\FunctionTok{c}\NormalTok{(}\StringTok{"f"}\NormalTok{, }\StringTok{"m"}\NormalTok{), }\AttributeTok{size =}\NormalTok{ n, }\AttributeTok{replace =} \ConstantTok{TRUE}\NormalTok{)}
\NormalTok{) }
  
\CommentTok{\# type I interval censored survival data:}
\NormalTok{simdata1 }\OtherTok{\textless{}{-}}\NormalTok{ covariates }\SpecialCharTok{|\textgreater{}}
  \FunctionTok{mutate}\NormalTok{(}
    \AttributeTok{time =} \FunctionTok{rphreg}\NormalTok{(}\AttributeTok{u =} \FunctionTok{runif}\NormalTok{(n), }\SpecialCharTok{\textasciitilde{}}\NormalTok{ age}\SpecialCharTok{+}\NormalTok{sex, }\AttributeTok{beta =} \FunctionTok{c}\NormalTok{(}\DecValTok{1}\NormalTok{, }\FloatTok{0.5}\NormalTok{), }\AttributeTok{dist =} \StringTok{"exp"}\NormalTok{, }\AttributeTok{rate =} \DecValTok{1}\NormalTok{),}
    \AttributeTok{tau =} \FunctionTok{rweibull}\NormalTok{(n, }\AttributeTok{scale =} \DecValTok{2}\NormalTok{, }\AttributeTok{shape =} \FloatTok{1.5}\NormalTok{),}
    \FunctionTok{rinterval}\NormalTok{(time, tau, }\AttributeTok{type =} \StringTok{"I"}\NormalTok{)}
\NormalTok{  )}
\FunctionTok{glimpse}\NormalTok{(simdata1)}
\end{Highlighting}
\end{Shaded}

\begin{verbatim}
## Rows: 300
## Columns: 6
## $ age   <dbl> 1.34592454, 0.99527131, 0.54622688, -1.91272392, 1.92128431, 1.3~
## $ sex   <chr> "m", "f", "m", "m", "m", "f", "m", "f", "f", "m", "m", "f", "f",~
## $ time  <dbl> 0.05387373, 0.05326879, 0.16546309, 2.17932784, 0.04415598, 0.01~
## $ tau   <dbl> 0.8276881, 1.4807894, 3.2631929, 3.3893988, 2.7803610, 0.4764430~
## $ left  <dbl> 0.0000000, 0.0000000, 0.0000000, 0.0000000, 0.0000000, 0.0000000~
## $ right <dbl> 0.8276881, 1.4807894, 3.2631929, 3.3893988, 2.7803610, 0.4764430~
\end{verbatim}

\begin{Shaded}
\begin{Highlighting}[]
\NormalTok{fit1 }\OtherTok{\textless{}{-}} \FunctionTok{ic\_par}\NormalTok{(}
  \FunctionTok{cbind}\NormalTok{(left, right) }\SpecialCharTok{\textasciitilde{}}\NormalTok{ age}\SpecialCharTok{+}\NormalTok{sex, }\AttributeTok{data =}\NormalTok{ simdata1,}
  \AttributeTok{model =} \StringTok{"ph"}\NormalTok{, }\AttributeTok{dist =} \StringTok{"exponential"}
\NormalTok{)}
\FunctionTok{summary}\NormalTok{(fit1)}
\end{Highlighting}
\end{Shaded}

\begin{verbatim}
## 
## Model:  Cox PH
## Dependency structure assumed: Independence
## Baseline:  exponential 
## Call: ic_par(formula = cbind(left, right) ~ age + sex, data = simdata1, 
##     model = "ph", dist = "exponential")
## 
##           Estimate Exp(Est) Std.Error z-value         p
## log_scale  -0.3877   0.6786    0.1038  -3.736 1.871e-04
## age         0.9186   2.5060    0.1241   7.402 1.341e-13
## sexm        0.2245   1.2520    0.1965   1.142 2.534e-01
## 
## final llk =  -94.91423 
## Iterations =  9
\end{verbatim}

The next example illustrates how type II interval-censored survival data
can be generated using the function \code{rinterval()}. In that example,
we assume that patients are expected to return to the physician at
certain pre-specified times, with attendancy probability equals to 0.7.

\begin{Shaded}
\begin{Highlighting}[]
\CommentTok{\# type II interval censored survival data:}
\NormalTok{simdata2 }\OtherTok{\textless{}{-}}\NormalTok{ covariates }\SpecialCharTok{|\textgreater{}}
  \FunctionTok{mutate}\NormalTok{(}
    \AttributeTok{time =} \FunctionTok{raftreg}\NormalTok{(}\AttributeTok{u =} \FunctionTok{runif}\NormalTok{(n), }\SpecialCharTok{\textasciitilde{}}\NormalTok{ age}\SpecialCharTok{+}\NormalTok{sex, }\AttributeTok{beta =} \FunctionTok{c}\NormalTok{(}\DecValTok{1}\NormalTok{, }\FloatTok{0.5}\NormalTok{), }\AttributeTok{dist =} \StringTok{"exp"}\NormalTok{, }\AttributeTok{rate =} \DecValTok{1}\NormalTok{),}
    \FunctionTok{rinterval}\NormalTok{(time, }\AttributeTok{tau =} \FunctionTok{seq}\NormalTok{(}\DecValTok{0}\NormalTok{, }\DecValTok{5}\NormalTok{, }\AttributeTok{by =} \DecValTok{1}\NormalTok{), }\AttributeTok{type =} \StringTok{"II"}\NormalTok{, }\AttributeTok{prob =} \FloatTok{0.7}\NormalTok{)}
\NormalTok{  )}
\FunctionTok{glimpse}\NormalTok{(simdata2)}
\end{Highlighting}
\end{Shaded}

\begin{verbatim}
## Rows: 300
## Columns: 5
## $ age   <dbl> 1.34592454, 0.99527131, 0.54622688, -1.91272392, 1.92128431, 1.3~
## $ sex   <chr> "m", "f", "m", "m", "m", "f", "m", "f", "f", "m", "m", "f", "f",~
## $ time  <dbl> 2.583140967, 2.152683999, 0.350258378, 0.045283358, 20.802515190~
## $ left  <dbl> 2, 1, 0, 0, 5, 2, 0, 1, 0, 1, 0, 0, 1, 2, 2, 0, 0, 1, 3, 5, 2, 3~
## $ right <dbl> 3, Inf, 1, 1, Inf, 3, 1, 2, 1, 3, 1, 1, 2, 4, 5, 2, 1, 2, Inf, I~
\end{verbatim}

\begin{Shaded}
\begin{Highlighting}[]
\NormalTok{fit2 }\OtherTok{\textless{}{-}} \FunctionTok{ic\_par}\NormalTok{(}
  \FunctionTok{cbind}\NormalTok{(left, right) }\SpecialCharTok{\textasciitilde{}}\NormalTok{ age}\SpecialCharTok{+}\NormalTok{sex, }\AttributeTok{data =}\NormalTok{ simdata2,}
  \AttributeTok{model =} \StringTok{"aft"}\NormalTok{, }\AttributeTok{dist =} \StringTok{"exponential"}
\NormalTok{)}
\FunctionTok{summary}\NormalTok{(fit2)}
\end{Highlighting}
\end{Shaded}

\begin{verbatim}
## 
## Model:  Accelerated Failure Time
## Dependency structure assumed: Independence
## Baseline:  exponential 
## Call: ic_par(formula = cbind(left, right) ~ age + sex, data = simdata2, 
##     model = "aft", dist = "exponential")
## 
##           Estimate Exp(Est) Std.Error z-value         p
## log_scale   0.3638    1.439    0.0716   5.081 3.747e-07
## age         1.0450    2.842    0.0908  11.500 0.000e+00
## sexm        0.4921    1.636    0.1480   3.326 8.825e-04
## 
## final llk =  -266.4422 
## Iterations =  7
\end{verbatim}

\subsection{Multivariate survival data, dependent censoring and
competing
risks}\label{multivariate-survival-data-dependent-censoring-and-competing-risks}

Copulas provide a way to construct multivariate distributions based upon
independent marginal distributions \citep{RBookCopula}. A multivariate
survival function can be constructed by the specification of independent
marginal survival models and an appropriated copula function.

Let \(C_{\theta}(u_{1}, \cdots, u_{d})\) a d-variated copula. Then, it
follows that the joint survival function of \((T_{1}, \cdots, T_{d})\)
can be expressed as:

\begin{equation}
  S(t_{1}, \cdots, t_{d}) = C_{\theta}\left(S_{1}(t_{1}|\boldsymbol{\Theta}_{1}, \mathbf{x}_{1}), \cdots, S_{d}(t_{d}|\boldsymbol{\Theta}_{d}, \mathbf{x}_{d})\right).
\end{equation}

To generate a random sample of \(\mathbf{T} = (T_{1}, \cdots,  T_{d})\)
it is sufficient to generate a random sample from
\((U_{1}, \cdots, U_{d}) \sim C_{\theta}(u_{1}, \cdots, u_{d})\) and
then take
\(T_{i} = S_{i}^{-1}(U_{i}|\boldsymbol{\Theta}_{i}, \mathbf{x}_{i})\).

\begin{Shaded}
\begin{Highlighting}[]
\FunctionTok{library}\NormalTok{(copula)}
\FunctionTok{library}\NormalTok{(rsurv)}

\FunctionTok{set.seed}\NormalTok{(}\DecValTok{1234567890}\NormalTok{)}

\NormalTok{n }\OtherTok{\textless{}{-}} \DecValTok{1000}   \CommentTok{\# sample size}
\NormalTok{tau }\OtherTok{\textless{}{-}} \FloatTok{0.5}  \CommentTok{\# Kendalls tau correlation}
\NormalTok{theta }\OtherTok{\textless{}{-}} \FunctionTok{iTau}\NormalTok{(}\AttributeTok{copula =} \FunctionTok{claytonCopula}\NormalTok{(), }\AttributeTok{tau=}\NormalTok{tau)}
\NormalTok{clayton }\OtherTok{\textless{}{-}} \FunctionTok{claytonCopula}\NormalTok{(}\AttributeTok{param =}\NormalTok{ theta, }\AttributeTok{dim =} \DecValTok{2}\NormalTok{)}
\NormalTok{u }\OtherTok{\textless{}{-}} \FunctionTok{rCopula}\NormalTok{(clayton, }\AttributeTok{n =}\NormalTok{ n)}

\CommentTok{\# simulating the failure times:}
\NormalTok{simdata }\OtherTok{\textless{}{-}} \FunctionTok{data.frame}\NormalTok{(}
  \AttributeTok{age =} \FunctionTok{rnorm}\NormalTok{(n),}
  \AttributeTok{sex =} \FunctionTok{sample}\NormalTok{(}\FunctionTok{c}\NormalTok{(}\StringTok{"f"}\NormalTok{, }\StringTok{"m"}\NormalTok{), }\AttributeTok{size =}\NormalTok{ n, }\AttributeTok{replace =} \ConstantTok{TRUE}\NormalTok{)}
\NormalTok{) }\SpecialCharTok{|\textgreater{}}
  \FunctionTok{mutate}\NormalTok{(}
    \AttributeTok{t1 =} \FunctionTok{rphreg}\NormalTok{(}\AttributeTok{u =}\NormalTok{ u[,}\DecValTok{1}\NormalTok{], }\SpecialCharTok{\textasciitilde{}}\NormalTok{ age }\SpecialCharTok{+}\NormalTok{ sex, }\AttributeTok{beta =} \FunctionTok{c}\NormalTok{(}\DecValTok{1}\NormalTok{, }\FloatTok{1.2}\NormalTok{), }\AttributeTok{dist =} \StringTok{"exp"}\NormalTok{, }\AttributeTok{rate =} \DecValTok{2}\NormalTok{),}
    \AttributeTok{t2 =} \FunctionTok{rporeg}\NormalTok{(}\AttributeTok{u =}\NormalTok{ u[,}\DecValTok{2}\NormalTok{], }\SpecialCharTok{\textasciitilde{}}\NormalTok{ age }\SpecialCharTok{+}\NormalTok{ sex, }\AttributeTok{beta =} \FunctionTok{c}\NormalTok{(}\FloatTok{0.8}\NormalTok{, }\FloatTok{1.1}\NormalTok{), }\AttributeTok{dist =} \StringTok{"exp"}\NormalTok{, }\AttributeTok{rate =} \DecValTok{1}\NormalTok{),}
\NormalTok{  )}
\FunctionTok{glimpse}\NormalTok{(simdata)}
\end{Highlighting}
\end{Shaded}

\begin{verbatim}
## Rows: 1,000
## Columns: 4
## $ age <dbl> -0.17469076, -0.15529194, -1.03781319, -1.56364665, 1.03961867, -0~
## $ sex <chr> "m", "m", "f", "f", "f", "m", "m", "m", "m", "f", "f", "f", "f", "~
## $ t1  <dbl> 0.016749142, 0.167458832, 0.245770610, 0.695707186, 0.061162858, 0~
## $ t2  <dbl> 0.084588793, 0.583402263, 0.175592119, 0.288876906, 0.298716691, 0~
\end{verbatim}

\begin{Shaded}
\begin{Highlighting}[]
\CommentTok{\# checking out the correlation:}
\NormalTok{simdata }\SpecialCharTok{|\textgreater{}}
  \FunctionTok{select}\NormalTok{(t1, t2) }\SpecialCharTok{|\textgreater{}}
  \FunctionTok{cor}\NormalTok{(}\AttributeTok{method =} \StringTok{"kendall"}\NormalTok{)}
\end{Highlighting}
\end{Shaded}

\begin{verbatim}
##           t1        t2
## t1 1.0000000 0.5306627
## t2 0.5306627 1.0000000
\end{verbatim}

\begin{Shaded}
\begin{Highlighting}[]
\CommentTok{\# adding (right) censoring:}
\NormalTok{simdata1 }\OtherTok{\textless{}{-}}\NormalTok{ simdata }\SpecialCharTok{|\textgreater{}}
  \FunctionTok{mutate}\NormalTok{(}
    \AttributeTok{c =} \FunctionTok{runif}\NormalTok{(n, }\DecValTok{0}\NormalTok{, }\DecValTok{5}\NormalTok{) }\CommentTok{\# random censoring}
\NormalTok{  ) }\SpecialCharTok{|\textgreater{}}
  \FunctionTok{rowwise}\NormalTok{() }\SpecialCharTok{|\textgreater{}}
  \FunctionTok{mutate}\NormalTok{(}
    \AttributeTok{t1 =} \FunctionTok{min}\NormalTok{(t1, c),}
    \AttributeTok{t2 =} \FunctionTok{min}\NormalTok{(t2, c),}
    \AttributeTok{status1 =} \FunctionTok{as.numeric}\NormalTok{(t1 }\SpecialCharTok{\textless{}}\NormalTok{ c),}
    \AttributeTok{status2 =} \FunctionTok{as.numeric}\NormalTok{(t2 }\SpecialCharTok{\textless{}}\NormalTok{ c),}
\NormalTok{  )}
\FunctionTok{glimpse}\NormalTok{(simdata1)}
\end{Highlighting}
\end{Shaded}

\begin{verbatim}
## Rows: 1,000
## Columns: 7
## Rowwise: 
## $ age     <dbl> -0.17469076, -0.15529194, -1.03781319, -1.56364665, 1.03961867~
## $ sex     <chr> "m", "m", "f", "f", "f", "m", "m", "m", "m", "f", "f", "f", "f~
## $ t1      <dbl> 0.016749142, 0.167458832, 0.245770610, 0.695707186, 0.06116285~
## $ t2      <dbl> 0.084588793, 0.583402263, 0.175592119, 0.288876906, 0.29871669~
## $ c       <dbl> 2.26609325, 4.57257662, 3.63517003, 2.02494996, 2.66336179, 2.~
## $ status1 <dbl> 1, 1, 1, 1, 1, 1, 1, 1, 1, 1, 1, 1, 1, 1, 1, 1, 1, 1, 1, 1, 0,~
## $ status2 <dbl> 1, 1, 1, 1, 1, 1, 1, 1, 1, 1, 1, 1, 1, 0, 1, 1, 1, 1, 1, 1, 0,~
\end{verbatim}

\begin{Shaded}
\begin{Highlighting}[]
\CommentTok{\# checking out the correlation:}
\NormalTok{simdata1 }\SpecialCharTok{|\textgreater{}}
  \FunctionTok{select}\NormalTok{(t1, t2) }\SpecialCharTok{|\textgreater{}}
  \FunctionTok{cor}\NormalTok{(}\AttributeTok{method =} \StringTok{"kendall"}\NormalTok{)}
\end{Highlighting}
\end{Shaded}

\begin{verbatim}
##           t1        t2
## t1 1.0000000 0.5121882
## t2 0.5121882 1.0000000
\end{verbatim}

In the next example we show how to simulate survival data with dependent
censoring. In that case, the observable survival time is given by
\(Y = \min\{T, C\, A\}\), where \(T\), \(C\), and \(A\) represents the
time to failure, the dependent censoring time, and the censoring time
due to random censoring.

\begin{Shaded}
\begin{Highlighting}[]
\NormalTok{simdata2 }\OtherTok{\textless{}{-}}\NormalTok{ simdata }\SpecialCharTok{|\textgreater{}}
  \FunctionTok{mutate}\NormalTok{(}
    \AttributeTok{a =} \FunctionTok{runif}\NormalTok{(n, }\DecValTok{0}\NormalTok{, }\DecValTok{5}\NormalTok{) }\CommentTok{\# random censoring}
\NormalTok{  ) }\SpecialCharTok{|\textgreater{}}
  \FunctionTok{rowwise}\NormalTok{() }\SpecialCharTok{|\textgreater{}}
  \FunctionTok{mutate}\NormalTok{(}
    \AttributeTok{y =} \FunctionTok{min}\NormalTok{(t1, t2, a),}
    \AttributeTok{status1 =} \FunctionTok{as.numeric}\NormalTok{(y }\SpecialCharTok{==}\NormalTok{ t1),}
    \AttributeTok{status2 =} \FunctionTok{as.numeric}\NormalTok{(y }\SpecialCharTok{==}\NormalTok{ t2),}
\NormalTok{  ) }\SpecialCharTok{|\textgreater{}}
  \FunctionTok{select}\NormalTok{(}\SpecialCharTok{{-}}\FunctionTok{c}\NormalTok{(t1, t2))}
\FunctionTok{glimpse}\NormalTok{(simdata2)}
\end{Highlighting}
\end{Shaded}

\begin{verbatim}
## Rows: 1,000
## Columns: 6
## Rowwise: 
## $ age     <dbl> -0.17469076, -0.15529194, -1.03781319, -1.56364665, 1.03961867~
## $ sex     <chr> "m", "m", "f", "f", "f", "m", "m", "m", "m", "f", "f", "f", "f~
## $ a       <dbl> 3.1886887, 3.6066255, 3.0834830, 0.8921447, 0.8049795, 0.96584~
## $ y       <dbl> 0.016749142, 0.167458832, 0.175592119, 0.288876906, 0.06116285~
## $ status1 <dbl> 1, 1, 0, 0, 1, 1, 1, 1, 0, 1, 1, 0, 1, 1, 1, 1, 1, 1, 1, 0, 1,~
## $ status2 <dbl> 0, 0, 1, 1, 0, 0, 0, 0, 1, 0, 0, 1, 0, 0, 0, 0, 0, 0, 0, 1, 0,~
\end{verbatim}

Notice that the example presented above can be viewed as a particular
case of competing risks survival data. In the next example, we consider
a scenario where elements are subjected to three competing causes of
failure.

\begin{Shaded}
\begin{Highlighting}[]
\NormalTok{n }\OtherTok{\textless{}{-}} \DecValTok{1000}
\NormalTok{tau }\OtherTok{\textless{}{-}} \FloatTok{0.5}
\NormalTok{theta }\OtherTok{\textless{}{-}} \FunctionTok{iTau}\NormalTok{(}\AttributeTok{copula =} \FunctionTok{gumbelCopula}\NormalTok{(), }\AttributeTok{tau=}\NormalTok{tau)}
\NormalTok{gumbel }\OtherTok{\textless{}{-}} \FunctionTok{gumbelCopula}\NormalTok{(}\AttributeTok{param =}\NormalTok{ theta, }\AttributeTok{dim =} \DecValTok{3}\NormalTok{)}
\NormalTok{u }\OtherTok{\textless{}{-}} \FunctionTok{rCopula}\NormalTok{(gumbel, }\AttributeTok{n =}\NormalTok{ n)}

\NormalTok{simdata }\OtherTok{\textless{}{-}} \FunctionTok{data.frame}\NormalTok{(}
  \AttributeTok{age =} \FunctionTok{rnorm}\NormalTok{(n),}
  \AttributeTok{sex =} \FunctionTok{sample}\NormalTok{(}\FunctionTok{c}\NormalTok{(}\StringTok{"f"}\NormalTok{, }\StringTok{"m"}\NormalTok{), }\AttributeTok{size =}\NormalTok{ n, }\AttributeTok{replace =} \ConstantTok{TRUE}\NormalTok{)}
\NormalTok{) }\SpecialCharTok{|\textgreater{}}
  \FunctionTok{mutate}\NormalTok{(}
    \AttributeTok{t1 =} \FunctionTok{rphreg}\NormalTok{(}\AttributeTok{u =}\NormalTok{ u[,}\DecValTok{1}\NormalTok{], }\SpecialCharTok{\textasciitilde{}}\NormalTok{ age }\SpecialCharTok{+}\NormalTok{ sex, }\AttributeTok{beta =} \FunctionTok{c}\NormalTok{(}\DecValTok{1}\NormalTok{, }\FloatTok{1.2}\NormalTok{), }\AttributeTok{dist =} \StringTok{"lnorm"}\NormalTok{, }\AttributeTok{meanlog =} \DecValTok{0}\NormalTok{, }\AttributeTok{sdlog =} \DecValTok{1}\NormalTok{),}
    \AttributeTok{t2 =} \FunctionTok{rphreg}\NormalTok{(}\AttributeTok{u =}\NormalTok{ u[,}\DecValTok{2}\NormalTok{], }\SpecialCharTok{\textasciitilde{}}\NormalTok{ age }\SpecialCharTok{+}\NormalTok{ sex, }\AttributeTok{beta =} \FunctionTok{c}\NormalTok{(}\FloatTok{0.8}\NormalTok{, }\FloatTok{1.1}\NormalTok{), }\AttributeTok{dist =} \StringTok{"exp"}\NormalTok{, }\AttributeTok{rate =} \DecValTok{1}\NormalTok{),}
    \AttributeTok{t3 =} \FunctionTok{rphreg}\NormalTok{(}\AttributeTok{u =}\NormalTok{ u[,}\DecValTok{3}\NormalTok{], }\SpecialCharTok{\textasciitilde{}}\NormalTok{ age }\SpecialCharTok{+}\NormalTok{ sex, }\AttributeTok{beta =} \FunctionTok{c}\NormalTok{(}\FloatTok{0.7}\NormalTok{, }\FloatTok{1.0}\NormalTok{), }\AttributeTok{dist =} \StringTok{"exp"}\NormalTok{, }\AttributeTok{rate =} \DecValTok{1}\NormalTok{),}
    \AttributeTok{a =} \FunctionTok{runif}\NormalTok{(n, }\DecValTok{0}\NormalTok{, }\DecValTok{5}\NormalTok{) }\CommentTok{\# random censoring}
\NormalTok{  ) }\SpecialCharTok{|\textgreater{}}
  \FunctionTok{rowwise}\NormalTok{() }\SpecialCharTok{|\textgreater{}}
  \FunctionTok{mutate}\NormalTok{(}
    \AttributeTok{y =} \FunctionTok{min}\NormalTok{(t1, t2, t3, a),}
    \AttributeTok{status1 =} \FunctionTok{as.numeric}\NormalTok{(y }\SpecialCharTok{==}\NormalTok{ t1),}
    \AttributeTok{status2 =} \FunctionTok{as.numeric}\NormalTok{(y }\SpecialCharTok{==}\NormalTok{ t2),}
    \AttributeTok{status3 =} \FunctionTok{as.numeric}\NormalTok{(y }\SpecialCharTok{==}\NormalTok{ t3),}
\NormalTok{  ) }\SpecialCharTok{|\textgreater{}}
  \FunctionTok{select}\NormalTok{(}\SpecialCharTok{{-}}\FunctionTok{c}\NormalTok{(t1, t2, t3))}
\FunctionTok{glimpse}\NormalTok{(simdata)}
\end{Highlighting}
\end{Shaded}

\begin{verbatim}
## Rows: 1,000
## Columns: 7
## Rowwise: 
## $ age     <dbl> -0.53451004, -1.48095715, -0.76892438, 0.43319025, 0.76357712,~
## $ sex     <chr> "m", "m", "m", "m", "m", "m", "m", "f", "f", "f", "f", "m", "f~
## $ a       <dbl> 2.17617891, 1.49961178, 4.65348449, 1.04816011, 1.12082428, 0.~
## $ y       <dbl> 0.18034466, 0.51001897, 0.07337358, 0.12327447, 0.13908667, 0.~
## $ status1 <dbl> 0, 1, 0, 0, 0, 0, 0, 0, 0, 0, 0, 0, 0, 0, 0, 1, 0, 0, 0, 0, 0,~
## $ status2 <dbl> 0, 0, 1, 1, 0, 0, 1, 0, 0, 1, 1, 0, 0, 1, 0, 0, 0, 0, 1, 1, 0,~
## $ status3 <dbl> 1, 0, 0, 0, 1, 0, 0, 0, 1, 0, 0, 1, 1, 0, 0, 0, 1, 1, 0, 0, 1,~
\end{verbatim}

\section{Conclusions}\label{sec-conclusions}

This paper has presented the new \proglang{R} package \pkg{rsurv}. The
proposed package is designed for general survival data simulation. It
combines a wide range of survival regression models, which includes AFT,
PH, PO, AH, YP and EH models, with an unlimited number of baseline
survival distributions. The package \pkg{rsurv} is suitable to
generating right, left and interval-censored survival data under
independent and dependent censoring mechanisms. The functions provided
in the package \pkg{rsurv} also allows the generation of survival data
with cure fraction, survival data with random effects (frailties), and
competing risks survival data. Another nice characteristic of the
package \pkg{rsurv} regards the fact that linear predictors are
specified through R formulas. This, together with the use of \pkg{dplyr}
verbs, makes the survival data simulation process much more flexible and
intuitive.

Future research involves the development of new functions for simulating
from regression models with time-dependent covariates, and time-varying
covariate effects. Further possible extensions include the
implementation of methods to simulate survival data from recurrent
evetns, and simulate data from truncaded survival data.

\section*{Appendix}\label{appendix}
\addcontentsline{toc}{section}{Appendix}

\subsection*{Proof of Theorem 1}\label{proof-of-theorem-1}
\addcontentsline{toc}{subsection}{Proof of Theorem 1}

We start the proof of Theorem 1 by the YP model given in Equation
\eqref{survYP}. For this step of the proof, define
\(\kappa_{1} = e^{\eta_{1}}\) and \(\kappa_{2} = e^{\eta_{2}}\).

\begin{itemize}
  \item[i)] As $S_{0}(t|\boldsymbol{\theta})$ does not depend on $\boldsymbol{\eta}$, we have that $a(t, \boldsymbol{\eta}) = t$. Using the fact that $R_{0}(t|\boldsymbol{\theta}) = \displaystyle \frac{1}{1+S_{0}(t|\boldsymbol{\theta})} = \frac{1}{1+s}$, it follows that
 
 \begin{equation}
S(t|\boldsymbol{\Theta}, \mathbf{x}) = \left[1+R_{0}(t|\boldsymbol{\theta})\frac{\kappa_{1}}{\kappa_{2}}\right]^{-\kappa_{2}} 
= \left[1+\frac{\kappa_{1}}{\kappa_{2}}\frac{1}{1+s}\right]^{-\kappa_{2}} 
= g(s, \boldsymbol{\eta}). \nonumber
\end{equation}

  \item[ii)] We have that

\begin{eqnarray*}
u = S(t|\boldsymbol{\Theta}, \mathbf{x}) = \left[1+\frac{\kappa_{1}}{\kappa_{2}}\frac{1}{1+s}\right]^{-\kappa_{2}} 
&\Leftrightarrow& 
v = \left[u^{-1/\kappa_{2}} - 1 \right]\frac{\kappa_{2}}{\kappa_{1}} = \frac{1}{1+s} \\
&\Leftrightarrow&
s = \frac{1}{1+v} = g^{-1}(u, \boldsymbol{\eta}).
\end{eqnarray*}

Therefore, since $a(t, \boldsymbol{\eta}) = t$, we have that

\begin{equation}
t = S_{0}^{-1}\left(\frac{1}{1+v}\right) = S_{0}^{-1}\left(g^{-1}(u)\right). \nonumber
\end{equation}

\end{itemize}

Now we present the proof of Theorem 1 for the EH model given in Equation
\eqref{survEH}.

\begin{itemize}

\item[i)] We have that $a(t, \boldsymbol{\eta}) = t/e^{\eta_{1}}$. It follows directly from Equation \eqref{survEH} that

\begin{eqnarray}
S(t|\boldsymbol{\Theta}, \mathbf{x}) &=& \exp\left\{-H_{0}\left(t/e^{\mathbf{x}\boldsymbol{\beta}}|\boldsymbol{\theta})e^{\mathbf{x}(\boldsymbol{\beta} + \boldsymbol{\phi}}\right)\right\} \nonumber \\
&=& S_{0}\left(t/e^{\mathbf{x}\boldsymbol{\beta}}|\boldsymbol{\theta}\right)^{e^{\mathbf{x}(\boldsymbol{\beta} + \boldsymbol{\phi})}} \nonumber \\
&=& s^{e^{\eta_{1}+\eta_{2}}} = g(s, \boldsymbol{\eta}).  \nonumber
\end{eqnarray}

\item[ii)] Note that

\begin{eqnarray*}
u = S(t|\boldsymbol{\Theta}, \mathbf{x}) &=& s^{e^{\eta_{1}+\eta_{2}}} \\
&\Leftrightarrow& 
s = u^{-e^{\eta_{1}+\eta_{2}}} = g^{-1}(u, \boldsymbol{\eta}).
\end{eqnarray*}

Thus,

\begin{eqnarray*}
  a(t) = S_{0}^{-1}(s) &=& S_{0}^{-1}\left(g^{-1}(u)\right) \Leftrightarrow   t = a^{-1}\left(S_{0}^{-1}\left(g^{-1}(u)\right)\right).
\end{eqnarray*}

\end{itemize}

\subsection*{Proof of Corollary 1}\label{proof-of-corollary-1}
\addcontentsline{toc}{subsection}{Proof of Corollary 1}

It follows immediately from Theorem 2 that:

\begin{eqnarray*}
S_{\xi}(t|\boldsymbol{\Theta}, \mathbf{x}) = A_{p}^{-1}\left(S_{pop}(t)\right).
\end{eqnarray*}

The rest of the proof follows from \eqref{t_cure} and the direct
application of Theorem 1.

\bibliographystyle{Chicago}
\bibliography{REFERENCES.bib}

\end{document}